# Redback/Black Widow Systems as progenitors of the highest neutron star masses and low-mass Black Holes

J. E. Horvath [1*], A. Bernardo[1], L.S. Rocha[1], R. Valentim[2], P.H.R.S. Moraes[1] and

M.G.B. de Avellar[2]

[1]*Universidade de São Paulo, Department of Astronomy IAG-USP. R. do Matão 1226, 05508-090, Cidade Universitária, São Paulo SP, Brazil;*
[2]*Departamento de Física, Instituto de Ciências Ambientais, Químicas e Farmacêuticas (ICAQF). R. São Nicolau 210, 09913-030, Diadema, SP, Brazil*

The long-standing problem of the maximum mass that can be achieved by these compact objects, with clear implications for the equation of state of matter above the nuclear saturation density (see [1] and references therein), is receiving a new twist with the latest data from binary systems and NS-NS merging events. Long ago, double neutron stars systems (DNS) data installed the idea of a "canonical" mass of $\sim 1.4 \, M_\odot$, but later work provided evidence for heavier objects with increasing degree of confidence, and it became clear that at least a second ``mass scale'' had to be present [2]. This second mass scale contains NSs born massive plus the set of substantially accreted objects.

On the other hand, the celebrated event widely explained as the fusion of a DNS GW170817, was readily interpreted as evidence for the formation of a transitory state, most likely a hypermassive NS (HMNS), or a supermassive NS (SMNS), eventually forming a black hole (BH), or even a stable NS (SNS) [4]. Simple physical assumptions were later used to connect this transient state to the maximum mass [3,4,5] of a static NS sequence.

The group of "spiders" ("black widow" systems like PSR1957 + 20 and their redback "cousins"), are strongly interacting relativistic systems harboring one pulsar. They have been modeled as the result of two unusual ingredients, not present in ordinary LMXBs: the back illumination onto the donor and the later ablation of the donor by the pulsar wind, showing an evolutionary connection between the two groups [6,7]. In fact the observation of high-masses in "spiders" stems from the accretion history of the systems:

starting from the formation of the NS (highly variable accretion rates $\leq 10^{-9} \, M_\odot$), the accretion times until the donor starts to enter the degenerate regime are *very long*, in the ballpark of $4 - 5 \, Gyr$ (Ref. [6], Fig. 2) and allow the growth of the NS mass even for moderate efficiencies of accretion ($\beta \geq 0.1$), as required for the systems to evolve towards the ``black widow'' region in the orbital-donor mass plane within a Hubble time.

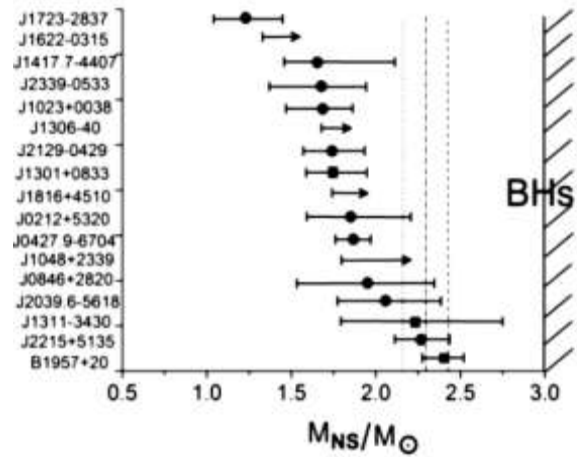

**Figure 1** The masses of 17 Redback/Black Widows NSs. The circles denote Redback systems and the squares Black Widow ones. Lower limits on the masses of J1622-0315, J1306-40, J1816+4510 and J1048+2339 are indicated with the arrows. The dotted vertical line is the maximum mass derived by Magalit and Metzger [4], the dashed-dotted line the upper limit of the range quoted by Ruiz, Shapiro and Tsokaros [5] and the dashed line the value of Ai, Gao and Zhang [3] for the SMNS formation case in the GW170817 merger (full data of the direct observations and references can be found in Özel and Freire [8]). The hatched region on the right marks the region of BHs, beyond the Rhoades-Ruffini limit, although objects pushed above $M_{max}$ will become BHs before the vertical line.

*Corresponding author (email: foton@iag.usp.br)

Fig. 1 depicts the NS masses of available "spider" systems. Our interpretation is that this is a distribution of accreted masses, starting with the NS masses at birth (*not* restricted to be $\sim 1.4 \, M_\odot$). The data is building a tension with some of the limits for the static maximum mass derived from the GW170817 merger event (vertical lines). Thus, higher statistics of "spider" systems has the potential of revealing directly the long-sought quantity $M_{max}$, if a statistically significant pile-up of masses happens at some limit $> 2 \, M_\odot$. Direct formation of $\geq 2 \, M_\odot$ NSs is certainly very problematic from the theoretical point of view.

Another important point is that some of the initially "spider" systems could have pushed their NSs over the TOV limit, forming a (small) number of low-mass black holes. This would be an alternative channel for BHs, not subject to constraints of gravitational collapse dynamics behind the "mass gap".

The observed spiders clearly cannot host a BH, but the prospects for detection of the "collapsed spiders", the ones that accreted sufficient mass for the NS to be "pushed" over the $M_{max}$ limit, can be enhanced by looking at accreting binary systems containing compact object candidates and companion stars complying with the features of a previous redback progenitor. One possible candidate is the system VLA J2130+12 [9], a low-luminosity source having a low-mass $0.1 - 0.2 \, M_\odot$ star as a companion and a very short orbital period of $1 - 2 \, h$. It should be noted that the common *assumption* of a "standard" BH mass of $10 \, M_\odot$ made in that work automatically rules out the possibility of identification of the object as a "collapsed spider", and a reanalysis is in order. There is a recent detection by LIGO (see https://gracedb.ligo.org/superevents/S200316bj/view/) which is being considered as a candidate for BHs ``filling the gap'', with a preliminary probability of 0.9957 in favor of this $3 - 5 \, M_\odot$ range . This would indicate the possibility of forming low-mass BHs in other channels, not only in the "collapsed spider" systems suggested here.